\documentclass[12pt]{article}

\usepackage{jinstpub}
                    
\usepackage{graphicx}
\usepackage{subcaption}

\usepackage{hyperref}

\title{\boldmath Fast Timing at Colliders: From Scintillator Hodoscopes to MicroPattern Arrays}

\author[a]{Sebastian~White}

\emailAdd{sebastian.white@cern.ch}

\affiliation[a]{  Laboratory of Instrumentation and Experimental Particle Physics (LIP), Portugal}

\abstract{ 
	We give a select overview of charged particle timing at colliders. Roughly speaking, three eras capture the evolution of timing arrays. While luminosity and multiplicity
dictate the detector technologies, there is also a need to review the various approaches to  signal capture and processing, and their role in changing conditions for calibration. 

	For the current era--the High-Luminosity LHC upgrades--the technique of dual-threshold timing can significantly simplify the crucial correction for amplitude walk.This correction often spans a time range an order of magnitude larger than the target resolution. Furthermore, because operating conditions vary in the high radiation environment, calibration can no longer be thought of as a one-time procedure. 
	
	The TOFHIR2C ASIC designed for a CMS upgrade can be configured to provide dual threshold timing. In this report, we use this capability to
demonstrate the power of this technique suitable for our new era.}
	

\begin{document}
\maketitle
\flushbottom

\section{Introduction}
\label{sec:Introduction}

	In many respects charged particle timing at colliders is part of a wider community emphasizing timing, principally for time-of-flight (TOF). Time-of-flight particle identification--computed from momentum and velocity measured in fixed target beams--naturally was adopted by the ISR and $e^+\text{--}e^-$ colliders. The common tool was the scintillator hodoscope with photomultiplier readout. During the same period, TOF techniques were similarly used for particle identification in mass spectrometry. In medical imaging, timing performance has a clear--though still distant--target of producing 3-D imaging for Positron Emission Tomography (PET) and continual improvements in time resolution are essential for narrowing the field of view. As we will see below, R\&D for PET has also led to an attractive option for the HL-LHC.
	
	Within this community,  several approaches have been developed for sensor readout to provide optimal control over calibrations for amplitude walk, thermal drift and 
other factors that can degrade the intrinsic resolution. We discuss these approaches below, and this paper will focus on the technique of dual-threshold timing which has--by chance--been available in one of the CMS timing ASICs.

	Although the wider timing community has largely advanced through improved sensor technologies, collider timing has followed a more complex trajectory. The principal driver of timing systems at HL-LHC has changed from TOF-based particle identification to pileup mitigation and the opportunities for calibration have changed significantly. Frequent re-calibration also becomes necessary when the radiation dose requires modification of operating points.
	
	Given the new HL-LHC environment, the commissioning of timing detectors should explore alternative signal processing modes available in the readout system. The need for frequent re-calibration and the absence of a common t$_0$ time reference (due to pileup) should drive the approach to signal processing. We will show how a system designed to record a leading-edge time and pulse amplitude benefits from the availability of a second discriminator to implement dual threshold timing, simplifying calibration. 
	
\section{Timing Methods}
\label{sec:Methods}

	It is not our purpose to review developments in sensor technology. For a recent overview see J.~Va'vra's review\cite{Jerry}. Likewise, time measurement has proceeded consistently from time to amplitude conversion to digital and with this progress ASICs typically provide a leading-edge timestamp and pulse-amplitude measurement using time-over-threshold (TOT).
So long as there are tools to calibrate amplitude walk vs. TOT this would seem reasonable. 

\subsection{Limitations imposed by the HL-LHC environment}

	At the HL-LHC, however, the familiar calibration tools become significantly harder to access. In-time pileup\footnote{In-time pileup is distinguished from random events falling outside the rf bunch envelope.} at the level of 140-200 events per bunch crossing produces a distribution of potential t$_0$'s with an rms spread of roughly 180 picoseconds. Even dedicated low-pileup runs would not solve the problem:  experiments designed to focus on p--p collisions do not have a built in detector to register the ``event time".

	These constraints motivate a re-examination of the techniques that best capture the true time of arrival of a signal. A standard overview is given in Ref.\cite{Leo}. The most common approaches include:
	
\begin{enumerate}
\item{Leading--edge timing with pulse amplitude measured for walk correction}
\item{Dual threshold timing}
\item{Constant Fraction Timing}
\item{Amplitude and Risetime Compensation}
\item{Waveform capture and fitting}
\end{enumerate}

	Among these, method 1) is the baseline approach in most HL-LHC upgrade detectors. Because the TOFHIR2C ASIC--designed for method 1)--also provides a second discriminator, we have been able to take data using method 2) and compare its performance directly. This report therefore focuses on \textbf{dual-threshold timing}.

\subsection{Historical Context}

	Dual threshold timing is not a new concept. Earlier collider experiments explored it as well. The SLAC Mark II  and III TOF systems
recorded two leading-edge thresholds as well as pulse amplitude. An early design for the ALICE TOF system employed dual threshold timing\cite{Pestov}.

\subsection{Why Dual-Threshold Timing is Attractive}

	Comparing to 1) there is the issue that recording pulse height (Q or TOT) does not really give anything like a derivation of amplitude walk\footnote{It can also lead to a degradation of the resolution when there is an unrelated late component to the signal such as Ref.\cite{picosec,Jerry1}}.It is simply a correlation that can be extracted from data in which time delay relative to a fixed time reference can be mapped out and parametrized- often using a higher order polynomial. If access to a stable reference time is removed--as is the case at the HL-LHC- there is no model from which to derive the walk correction.

	Dual-threshold timing, by contrast, provides a direct measurement of the pulse slope, which is the quantity that physically determines time-walk.
	
	A simple way to state the fundamental position of slope in these corrections is to express the corrected time at threshold as a Taylor series. 
	Consider a current pulse, I', which reaches an amplitude, I$_{threshold}$ at time, t. Expanding the pulse around the true arrival time, t$_0$,
	
\begin{equation}
I'_{thresh} [t]=I_0+ \frac{dI}{dt} \times [t-t_0]+ \frac{1}{2} \times\frac{d^2I}{dt^2}\times[t-t_0]^2.....
\end{equation}

	The constant $I_0$ absorbs an offset in the start of the pulse, which is expected to be common among channels. Furthermore, we assume that once the pulse reaches threshold we are in an essentially linear region so
	
\begin{equation}
[t-t_0]=d_0+d_1/\frac{dI}{dt}
\end{equation}
	
	To first order, the \textbf{walk correction is proportional to the inverse of the pulse slope}. The 2 constants differ in relative importance. 
\begin{itemize}
\item{d$_1$ has units of threshold (current). It was found in \cite{JINST} to vary from channel-to-channel in the system we will consider but in a way that could be extracted from calibration data.  The dual threshold times determine the denominator. In the following we will refer to d$_1$ as the amplitude walk coefficient (AWC).}
\item{We might think of t$_0$ as the actual time of arrival and then the parameter d$_0$ is an offset that makes this connection. For the present discussion we will focus on the second term--i.e. the walk correction--and consider determining the channel variations in d$_0$ as a last step, which compensates for clock distribution and other factors that are unlikely to change in an experiment.}
\end{itemize}
	
\section{Three periods of Timing Detectors at Colliders}

	It is convenient to separate the approaches to timing in Collider experiments into three distinct periods. For the five hadron colliders (ISR, Sp$\bar{p}$,Tevatron,RHIC,LHC) \footnote{I have done experiments at all 5, starting with my thesis in 2nd phase of ISR.} we will single out the LHC and HL-LHC for the 2nd and 3rd epoch and focus on the British-Scandinavian survey experiment at the ISR for the 1st epoch. For $e^+\text{--}e^-$ colliders, we focus on Mark II and III, although differences from Adone, Orsay, PETRA are worth pointing out. Later, and future, $e^+\text{--}e^-$ collider experiments tend to focus on alternatives to TOF for particle id.
	
\subsection{TOF Hodoscopes}

	At the ISR and Mark II, luminosity and multiplicities were low enough that a relatively small number of scintillation counters were sufficient to cover the TOF needs. Table I lists characteristics of Counters used in several experiments reproducing a Table by Atwood\cite{Atwood}. Key parameters include scintillator length, light yield and the measured time resolution. 
	
	With these first generation colliders a design consideration was to provide \textbf{beam counters}, recording the downstream debris from collisions. Time coincidence between these counters provided discrimination between real events and background and the beam-counter trigger could be used as an \textbf{event time reference} for TOF. At SLAC Mark II, III the machine RF provided an even more precise reference time for the collision and the somewhat better TOF resolution at SLAC compared to Orsay was attributed to SLAC's shorter bunch length. By contrast, the ISR operated as a DC machine so the beam counters had to do all of the work.

\begin{table}[!htp]
\centering
\begin{tabular}{|c|c|c|c|c|}
\hline
Counter& L(cm)& N$_{pe}$ & $\delta$t(picosec)& \\ \hline
Mark II& 350 & $40$ & $255$ & \\ \hline
DASP& 172 & $28$ & $212$ & \\ \hline
Binon & 25 & $28$ & $92$ & \\ \hline
Mark III & 300 & $120$ & $120$ & \\ \hline
\end{tabular}
\caption{Summary of Scintillation counters used in TOF systems -taken from Atwood 1982 SLAC summer Institute. Listed are the scintillator length, photoelectron yield and measured time resolution.}
\label{tab:walk2_coeffs}
\end{table}

	So for these experiments, a t$_0$ reference was a given and presumably a scheme following Method 1 was adequate for calibration. Nevertheless Mark II, III (and apparently PEP-2) used front end electronics that provided \textbf{two leading--edge thresholds} as well as pulse height measurement. A concern with organic scintillators at the time was significant attenuation over the length of a scintillator bar. This was addressed by choosing one or the other threshold depending on the pulse height range. Naturally, this system would also enable using both thresholds together for improved timing.  The Mark III Time of Flight System paper \cite{Mark} gives a very detailed description of their calibration procedure and, in spite of the fact that they didn't actually use the dual threshold feature for timing, there is an interesting summary, which we quote in full:
	
	``The upper threshold is used except when the pulse height is too small and the lower threshold value is
then used. No algorithm has yet been devised which, using both thresholds together, gives an improvement in
resolution for particle data, although laser tests suggest this may eventually be feasible"

	This is perhaps the only case of a collider detector with dual threshold capability, although it was originally considered for ALICE\cite{Pestov}.
	
	A final observation from the scintillator era concerns optimization efforts by the Mark II collaboration. These focused on attenuation and light yield (N$_{pe} $ in Table 1) and concluded that they were reaching a point of diminishing returns in time resolution- thicker scintillator being the only option.  In later development alternative approaches included Cerenkov radiators, faster transit time in the photosensors and also inorganic scintillators\cite{Inami,picosec,BTL}.

\subsection{ALICE}

	Fast timing in large acceptance collider experiments entered a new era with the commissioning of the \textbf{ALICE Time-of-Flight (TOF) detector} at the LHC. The Alice TOF 
system uses \textbf{Multigap Resistive Plate Chamber} technology. It consists of $\sim$150k channels and achieved a charged particle time resolution of \textbf{82 picoseconds} initially, and starting from LHC Run-2 improving to \textbf{52 picoseconds}\cite{ALICE2,ALICE} -largely due to improvements in the  corrections for \textbf{amplitude-dependent time--walk} of the leading--edge discriminator.

	The calibration effort required in ALICE was substantial and many aspects directly bear on the likely procedures to be employed in the new systems of ATLAS and CMS upgrades\cite{TDR}. However, ALICE benefited from operating in a \textbf{low-rate environment}. The ALICE TPC tracking system required an interaction rate so low that events only capture a \textbf{single interaction}, whereas the ATLAS and CMS experiments are designed to handle \textbf{140 to 200 p-p interactions per bunch crossing}, i.e., high pileup. 
	
	Although there is an extensive literature on \textbf{amplitude walk} and its correction\cite{Leo}, for the purpose of this report we will take the ALICE case as the canonical example of the traditional approach to calibration and compare it to an alternative that is enabled by the ASIC of the \textbf{CMS Barrel Timing Layer(BTL)} upgrade.
	A detailed discussion of the ALICE procedure and its lessons for ATLAS/CMS upgrades can be found in Ref. \cite{JINST}. Here we focus on a few relevant points. The timing calibration really consists of 2 separate steps:
	
\begin{itemize}
\item{ \textbf{walk calibration}, corrects for amplitude dependent shift of the leading edge time.}
\item{\textbf{channel-to-channel offset determination}, aligning all channels to a common reference.}
\end{itemize}

	One advantage of the traditional approach is that both steps are naturally incorporated in the same calibration workflow. In the approach presented here, we will put aside the issue of offset constants, which are comparatively easy to determine once the \textbf{walk correction} is under control.

\subsection{High Luminosity LHC Upgrades}

\begin{figure}[!htp]
\centering
\includegraphics[width=.6\textwidth]{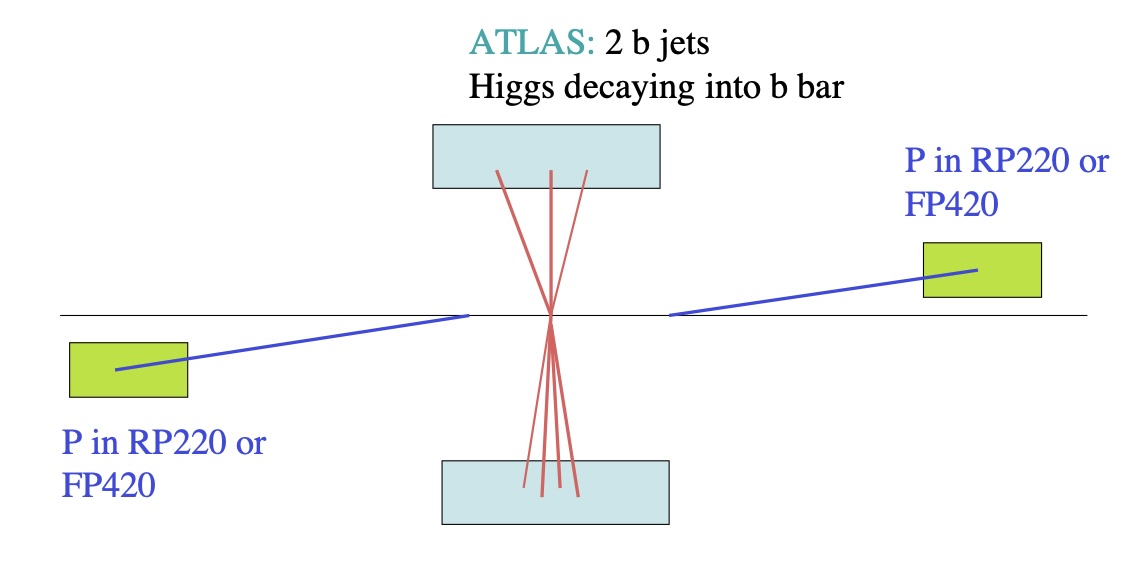}
\caption{ FP420 schematic for Central Exclusive Production of Higgs decaying to 2 B-jets. Forward protons are associated with z$_{vertex}$ using time of flight.}
\label{fig:cartoon}       
\end{figure}

	Fast timing entered LHC physics discussion well before the machine turned on.
	The \textbf{FP420 proto-collaboration} \cite{FP420}, emphasized a fundamentally different role for fast timing: enabling the association of \textbf{forward protons} with centrally produced physics objects in a high pileup environment.  Forward protons had played an important role at earlier hadron colliders--beginning with the ISR--enabling a broad diffractive physics program. At the LHC, however, exploiting this tool for potential new physics in \textbf{high-mass diffraction} required solving a new problem: associating forward protons with the correct central event vertex under conditions of extreme pileup.
	
	Previously, event components were assembled using the \textbf{tracking vertex}. But, forward protons detected at milliradian angles cannot be associated with a single vertex using tracking alone.
	
	The FP420 proposal included fast timing on the leading protons close to the beam. In  \textbf{Central Exclusive Production(CEP)}--diffractive production with two leading protons--the distance travelled by each proton from the vertex is given by the time difference since the path is a straight line and $\beta$=1.
	
\begin{equation}
z_{vertex}=\frac{c \times\delta t}{2}
\end{equation}

	The proton derived vertex could then be matched to physics objects in the central region using their tracking vertex.
	
	Although the FP420 was never realized, \textbf{timing of forward protons} has been a feature of diffractive studies in \textbf{TOTEM, ATLAS} and the \textbf{CMS Precision Proton Spectrometer(PPS)}\cite{PPS}.

\begin{figure}[!htp]
\centering
\includegraphics[width=.45\textwidth]{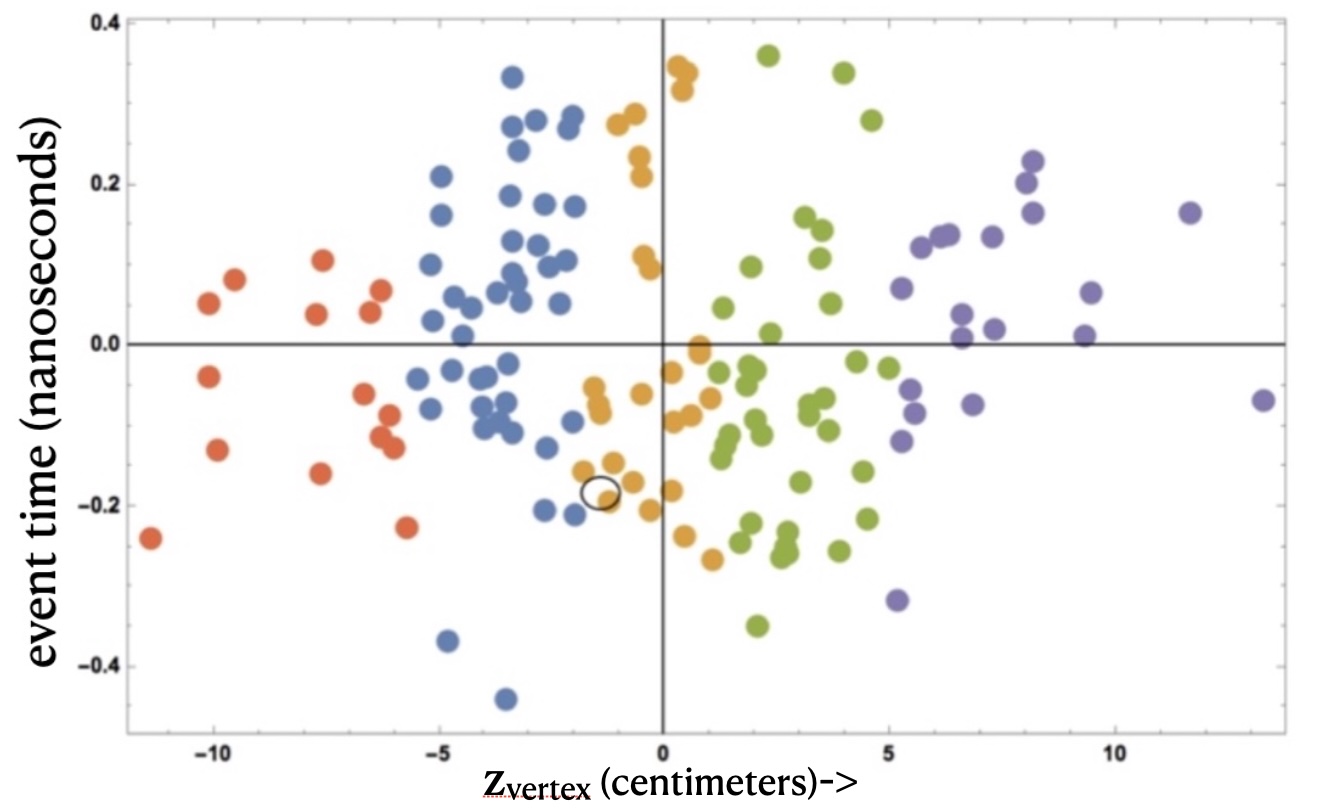}
\includegraphics[width=.45\textwidth]{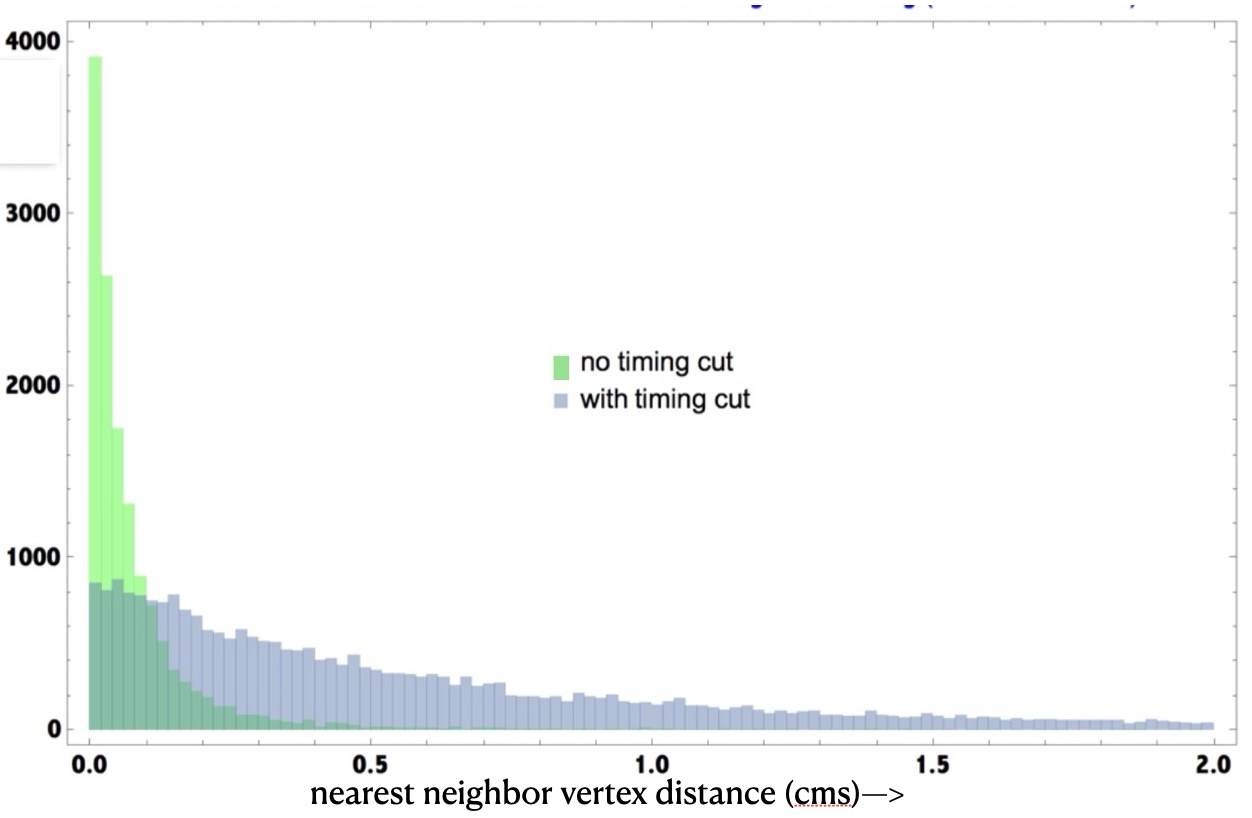}
\caption{ A typical HL-LHC bunch crossing interval giving the distribution of events in z$_{vertex}$ vs. event time with 140 events/crossing. Taken from 2007 preprint\cite{hermetic} by this author. The ellipse gives a hypothetical time and z resolution. Event colors 
are strictly arbitrary(left). Typical cleanup of nearest neighbor vertex candidates with time cuts (right).}
\label{fig:naive}       
\end{figure}

\subsubsection{Time-Space structure of HL-LHC Events}

	During FP420 proposal preparation, this author took the LHC design parameters (principally $\beta^*$, crossing angle and emittance) to model the \textbf{time and spacial distribution} of p-p event vertices in \textbf{ATLAS} or \textbf{CMS}\cite{hermetic}. The interest at the
time was principally for initial LHC luminosity but the HL-LHC nominal pileup (140 events/crossing) was also considered and an example is shown in Figure 2-left, illustrating the spread of events in z$_{vertex}$ and time. 

	What was intriguing at the time is that both ATLAS and CMS demonstrated \textbf{excellent time resolution} (e.g. for high energy electrons) in early testbeam studies of their EM calorimeters\cite{Campana,Rousseau}. Obviously, calorimeter time measurements in combination with leading proton times could lead to even better association with central objects. This is illustrated in Figure 2-right, which shows a typical improvement in vertex separation once one makes a timing cut. 

	Calorimeter timing has not yet been used in the analysis of forward proton data. However, this illustrated the power of timing for HL-LHC conditions to remove backgrounds due to mis-association of forward jets with central physics--particularly relevant for processes such as \textbf{vector boson fusion}.
	
\subsubsection{Towards Hermetic Timing at the HL-LHC}

	Over the past decade, ATLAS and CMS have been developing--and now have largely completed--upgrades to incorporate fast timing for pileup mitigation. For the CMS experiment the approach could be characterized as ``hermetic timing":
\begin{itemize}
\item{upgrades to the central electromagnetic calorimeter electronics}
\item{a new forward \textbf{High-Granularity Calorimeter(HGCAL)} with timing capability}
\item{extensive charged-particle timing coverage via the \textbf{MIP Timing Detector(MTD)} , consisting of the \textbf{Barrel Timing Layer(BTL)} and the \textbf{Endcap Timing Layer(ETL)}.}
\end{itemize}

	For the purposes of this report, we focus on the \textbf{BTL electronics} where we can directly explore the benefits of \textbf{dual-threshold timing} compared to the, more common, approach of \textbf{leading--edge timing and time-over-threshold} (ie Method 2) vs. 1) in the above section 2).

\section{Dual Threshold Timing with the TOFHIR2C ASIC}

	The CMS Barrel Timing Layer\cite{BTL} comprises approximately 165,000 LYSO scintillator bars (3.75 x 3.12 x 54.7 mm$^3$) readout by matching-area SiPMs on each end of the bar. The technology and the readout electronics benefit from previous history in PET imaging but the collaboration has invested heavily in an optimization of all aspects in order to deal with rates and radiation damage in the HL-LHC environment.

	The readout ASIC \textbf{TOFHIR2C} for the BTL is presented schematically in Figure 3. It features single delay line shaping in both \textbf{time} and \textbf{amplitude} branches for baseline restoration in the high-rate environment. Crucially, the ASIC includes a \textbf{second discriminator threshold} enabling the study and implementation of \textbf{ dual threshold timing}, which is the focus ot this section.
	
	In this report we will compare performance after calibration with an alternate approach to 1) (ie leading--edge timing and Q) benefiting from additional information available with the CMS BTL ASIC (TOFHIR2C)\cite{tahereh}. The new approach that we present here has already been demonstrated with bench test data from a pulsed UV laser emulating the signal in  BTL LYSO scintillators from charged particles\cite{JINST}. We refer the reader to this earlier report but simply recap a few aspects. The technique is also currently being validated using 180 GeV pion beam data\cite{note2}.
	
	The calibration method that we have proposed takes advantage of 2 available thresholds (thr$_1$ and thr$_2$) that can be independently set on the leading--edge in the TOFHIR2C ASIC. Using these thresholds, we can also extract the pulse-slope at threshold from their time difference\cite{Pestov,SLAC} . 
	
	We also find that, for the particular BTL/TOFHIR system under study, the measured pulse amplitude (Q) has a different correlation with the ``pulse slope at threshold" for every channel. Since the latter is the real determinant of walk, any calibration which is based instead on walk vs Q will clearly appear to be an empirical, brute force approach with no consistency among the 330 k channels of the BTL system.

\begin{figure}[!htp]
\centering
\includegraphics[width=.6\textwidth]{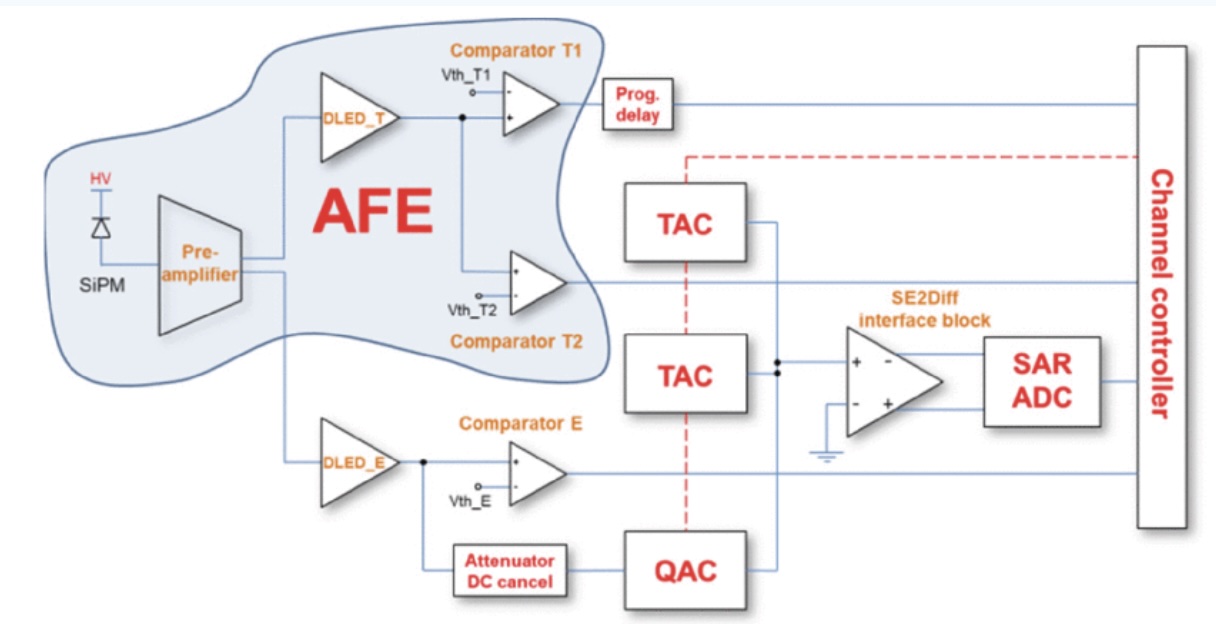}
\caption{ TOFHIR block diagram.}
\label{fig:cartoon}       
\end{figure}

\section{Calibration procedure}

	Prior to the present study, we used the LYSO/SiPM detector in a muon beam with custom trans-impedence amplifer and captured waveforms with a Lecroy Digital scope. This system exhibited perfect linearity in the sense that waveforms of different amplitude, once rescaled in amplitude, were identical. In this case the timing calibration reduced to a very simple procedure requiring only slope and amplitude measurement at one amplitude\cite{linear}.
	
	One final observation that is relevant for implementation of our approach is that the actual event-by-event jitter in slope accuracy is typically $\sim 13\%$ in the TOFHIR system whereas the pulse amplitude, Q, has higher precision-$\sim 3-5\%$. So there is a preliminary step to calibration based on ``pulse slope at threshold". To get the best accuracy we need to map Q to slope for each channel. This is easily done by accumulating several thousand hits in each BTL channel and fitting the trend which describes the mapping. As we will see, this simple calibration data set completely determines the walk correction for each channel.

\begin{figure}[!htp]
\centering
\includegraphics[width=.45\textwidth]{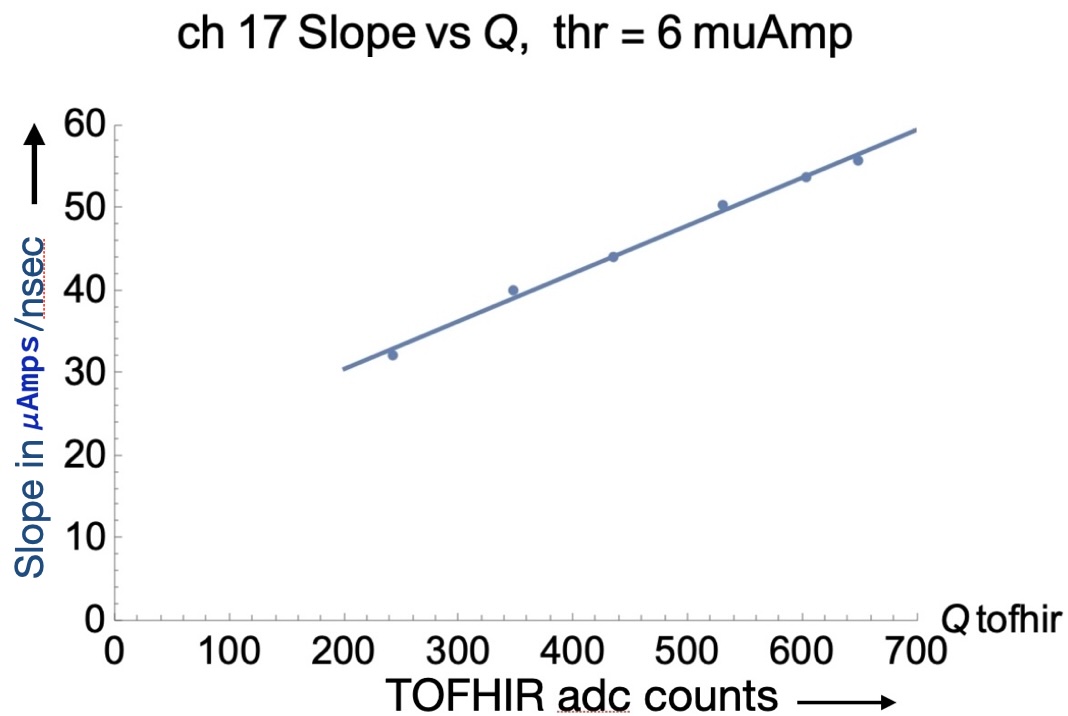}
\includegraphics[width=.40\textwidth]{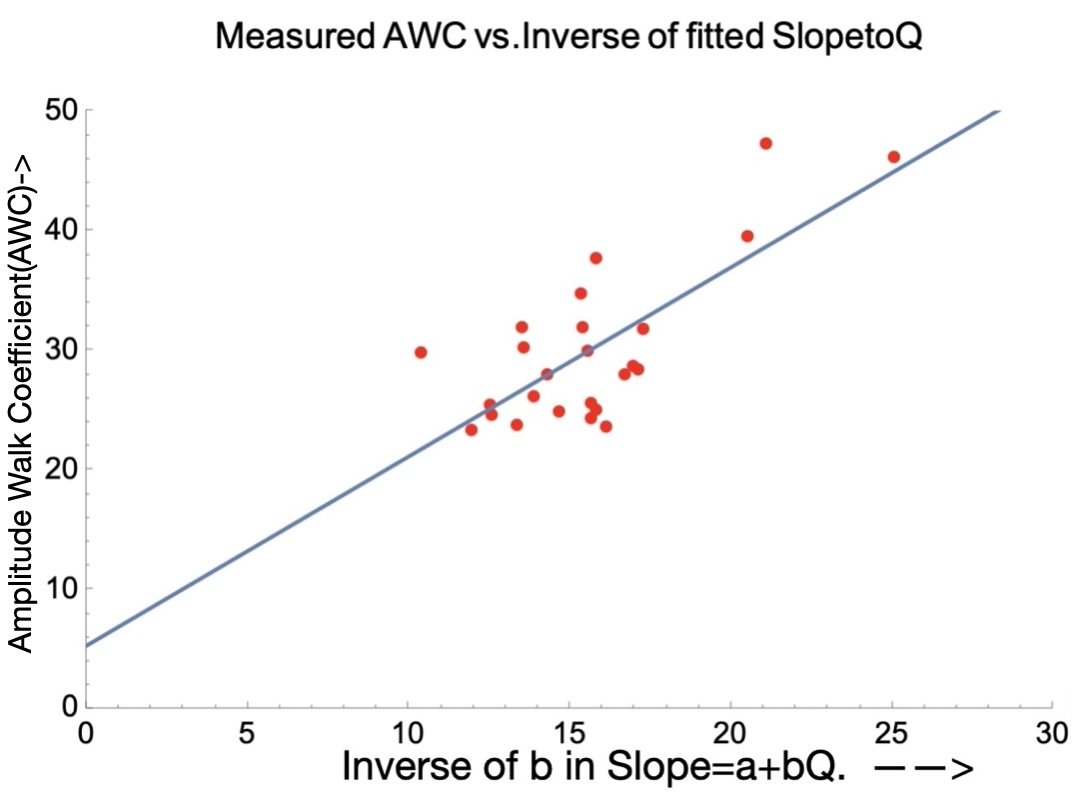}
\caption{ Slope vs q for a typical channel (left). The channel-to-channel spread in AWC values is correlated with a variation in Slope-to-Q ratio 
captured in calibration data- i.e. the b parameter derived from the left figure. Using this correlation we reduce the uncertainty on AWC to $\sim13\%$ (right).}
\label{fig:naive}       
\end{figure}

	The basic calibration for dual threshold timing consists in accumulating a look-up table or, more suitably, a fit as illustrated in Figure 4. Every channel of the TOFHIR2C has, more or less, different trend in Slope vs. Q as depicted here. Nevertheless this mapping is the determinant allowing a Q measurement to be inserted in eqn.2.2. What we have verified in Ref.\cite{JINST}, and in subsequent beam exposures, is that the linear form of eqn 2.2 continues to hold. This is already an interesting simplification since we now have a functional form and are dealing with a single parameter for full calibration of a 330k channel system.
	
\subsection{Amplitude Walk Coefficient Determination and Trimming}

	The lookup table or simple functional form as extracted in Figure 4 requires a calibration data set which is clearly simpler than the more sophisticated one used for the ALICE approach. We are simply accumulating data from the TOFHIR2C system where minimal cuts can assure that it is populated primarily with particle interactions in the LYSO crystals.
Typically a few thousand hits are sufficient, whether from early HL-LHC data or even cosmic rays.

	Given our experience with the laser tests and test beam data, we can be reasonably confident that the simple linear dependence on inverse slope will hold for each channel and the information accumulated with this simple calibration data set will serve as a first step be used to confirm this.
	
	What still remains to be done is to use the calibration data set to trim the channel-to-channel variations in the AWC. We, of course, now simply find empirically the individual channel
values for the AWC (i.e. d$_1$ in eqn. 2,.2) by comparing to slope dependence in test data where a t$_0$ reference was available (the laser time in the bench test and an MCP PMT\cite{note1}, with
$\leq 10$ picosecond jitter in the test beam\cite{MCP}). Having done this for our current sample, we find an rms spread in AWC values of $\leq 20\%$. This was perhaps sufficient for
an initial starting point to walk calibration but, in order to limit the walk contribution so that it is not affecting resolution, the goal should be of order $10\%$ spread.

	Up to now, analysis of the calibration data set, has led to tools that trim the channel-to-channel spread to roughly $13\%$\cite{JINST}. An example of this correction is given in Figure 5. It is likely that further work will reduce this spread to the needed level. There is now a large test beam data set with which to explore the validity of this approach.
	
	We should emphasize that the approach of this paper has been to establish a robust procedure for determining the amplitude walk coefficient (AWC or d$_1$ in eqn. 2.2). The starting position for this approach is that the remaining piece of the calibration--the channel-to-channel offset constants (d$_0$ in eqn. 2.2)-- should be comparatively simple to obtain once the walk variations are removed. The benefit of having a complete data set with a known t$_0$ is, of course, that both d$_0$ and d$_1$ could be obtained from these data.
	
	The other critical point about the HL-LHC environment is the impact of radiation damage which, for several systems, will lead to changes in operating point and therefore recalibration of walk. Time dependent variations in the d$_0$ offsets are less likely to be needed in these frequent re-calibrations.

\begin{figure}[!htp]
\centering
\includegraphics[width=.95\textwidth]{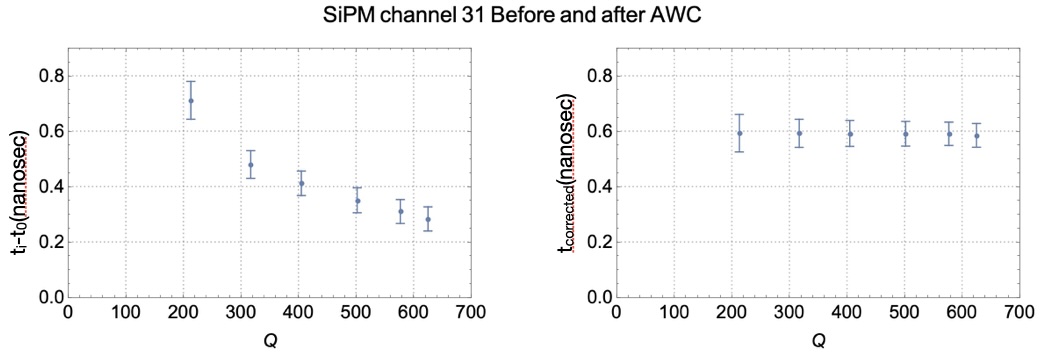}
\caption{ The procedure described in this paper removes the walk observed in the time at threshold (left) illustrated by applying eqn 2.2 to a typical channel (right). }
\label{fig:afterawc}       
\end{figure}

\section{Summary}

	There are a number of aspects of the new HL-LHC era in fast timing which present new challenges for timing detector calibration. We find it convenient to factorize two aspects of calibrating timing detector channels. Channel dependent time offsets (due to signal propagation, clock distribution, etc.) are less likely to depend on variations in detector gain and settings
over the detector lifetime. In any case they will be easier to measure once the amplitude walk variations are removed.

	Dual-threshold timing is an attractive approach to capturing the second term (d$_1$) since it can be used to capture the slope at threshold, which is physically determining amplitude walk. It removes, for example, the dependence on sensor gain or variations in electronic channel response. Our study demonstrated that, even for a system (BTL TOFHIR2C) that deviates for strict linearity(i.e. pulse shapes don't strictly scale with amplitude), the simple slope dependence of eqn. 2.2 holds.
	
	The simplification obtained by establishing this functional form is already significant. Clearly there is an obvious next step. If we can trim the spread in AWC values to a level closer to $13\%$ then we will have a \textbf{final} walk calibration rather than a good \textbf{initial} one.
	
	Beyond the particular upgrade project discussed here, it seems well worthwhile for other detector projects entering this new environment to review alternate approaches to calibrating 
for HL-LHC commissioning, still a few years off.

\section{Acknowledgement}

	I would like to acknowledge Alessio Boletti and Licheng Zhang who contributed to various stages of the analysis. Also Jerry Va'vra for useful discussions.
This work was partially supported, through Fermilab, by the US-CMS upgrade program. 

\end{document}